\documentclass[prb,aps,twocolumn,showpacs]{revtex4}

\usepackage{epsfig}
\usepackage{bm}

\newcommand{\beq}{\begin{equation}}
\newcommand{\eeq}{\end{equation}}
\newcommand{\bea}{\begin{eqnarray}}
\newcommand{\eea}{\end{eqnarray}}

\def\material{{Cs$_2$CuCl$_4$\ }}

\def\tit#1#2#3#4#5{{#1}{\bf #2}, #3 (#4)}

\def\npb{Nucl.\ Phys.\ B\ }
\def\plb{Phys.\ Lett.\ B\ }
\def\prl{Phys.\ Rev.\ Lett.\ }
\def\prb{Phys.\ Rev.\ B\ }
\def\jpco{J.\ Phys.\ Cond.\ Mat.\ }
\def\ijmpb{Int.\ J. Mod.\ Phys.\ B\ }
\def\pmag{Philos.\ Mag.\ }
\def\sci{Science\ }

\def\etal{{\em et al.}}
\def\vecb#1{{\bf #1}}

\begin{document}

\title{Ordering in \material: Is there a proximate spin liquid?}

\author{S. V. Isakov$^1$, T. Senthil$^{2,3}$,  and Yong Baek Kim,$^1$}

\affiliation{$^1$Department of Physics, University of Toronto,
Toronto, Ontario, Canada M5S 1A7\\
$^2$ Center for Condensed Matter Theory, Indian Institute of Science, Bangalore, India 560012\\
$^3$ Department of Physics, Massachusetts Institute of Technology, Cambridge, Massachusetts 02139}

\begin{abstract}
The layered spiral magnet \material displays several interesting properties 
that have been suggested as evidence of proximity to a two dimensional quantum spin liquid. 
In this paper we study a concrete version of this proposal and suggest 
experiments that can potentially confirm it.  
We study universal critical properties of two-dimensional frustrated
quantum magnets near the quantum phase transition between a spiral magnetic state 
and a spin liquid state with gapped bosonic spinons in the framework of the $O(4)$ invariant critical
theory proposed earlier\cite{chubukov}. Direct numerical 
calculation of the anomalous exponent in spin correlations shows 
that the critical scattering has broad
continua qualitatively similar to experiment. More remarkably we show that the enlarged $O(4)$ symmetry leads 
to the same slow power law decay for the vector spin chirality and the Neel correlations. We 
show how this may be observed through polarized neutron scattering experiments. A number of 
other less dramatic consequences of the critical theory are outlined as well.

\end{abstract}


\maketitle

\section{Introduction}

Despite the great amount of progress made in our theoretical 
understanding of spin liquid phases\cite{frustreview,fazekasanderson, anderson} of quantum magnets, 
there are to date no convincing experimental examples of such phases in spatial dimension $d \geq 2$. A particularly 
promising candidate - much discussed recently - is the material $Cs_2CuCl_4$. This is a layered Mott insulator 
whose spin physics is accurately modelled by a spin-$1/2$ Heisenberg antiferromagnet on an anisotropic 
triangular lattice. Neutron scattering measurements have revealed several unusual properties of the
spin excitations\cite{coldeaneu} which have motivated intensive theoretical research works
\cite{weihong,chungln,bocquet,chungsp,veillette}. 
There is incommensurate spiral magnetic order at low temperatures 
(below 0.62K). The low energy 
excitations of this ordered state are spin waves. However spin wave theory fails to quantitatively 
describe the spin wave line width seen in inelastic neutron scattering. In particular there are long tails 
that extend to reasonably high frequencies in the inelastic scattering spectrum. This structure may be understood as 
indicative of a broad continuum of excited states. The continuum survives upon heating 
above the magnetic phase transition (though the spin wave itself does not). In addition the phase diagram in an 
external magnetic field shows several interesting features.

The broad continuum is reminiscent of spinon excitations in one-dimensional Heisenberg antiferromagnetic
chain. However the continuum disperses in both spatial directions so that a two dimensional description
of the physics may be more appropriate. 
A number of workers have therefore suggested interpretations of the neutron data in terms of 
fractional spin (spinon) excitations in two dimensions
into which the magnons decay. In particular Coldea {\it et al.}\cite{coldeaneu} suggested that 
the material may be viewed as being close to a quantum phase transition between a spiral Neel state and a spin liquid state. 
The purpose of the present paper is to explore this possibility in greater detail  
and to suggest concrete experimental tests
to confirm (or rule out) this proposal. 

We begin with some general considerations. It is by now established that a number of different kinds of spin 
liquid phases are theoretically possible in two dimensional quantum magnets. We will restrict attention to a 
particular kind of spin liquid phase whose excitations consist of {\em bosonic} spin-$1/2$ spinons with a full 
spin gap. In addition there are gapped $Z_2$ vortices (visons) which act as sources of $\pi$ flux for the spinons. 
As shown many years ago in Ref.~\onlinecite{chubukov} such a spin liquid phase admits a direct second order 
transition to the spiral Neel state which is simply driven by 
condensation of the bosonic spinons. Specifically the transition was argued to be in the universality class 
of the classical $O(4)$ fixed point in three dimensions. 
Here we point out several remarkable consequences of this theory which may be used 
to test its applicability to \material or other materials. Perhaps most interestingly we show that the 
large extra $O(4)$ symmetry that emerges at the critical fixed point unifies seemingly different 
competing orders. The Neel vector correlations have the same slow power law decay as the vector spin chirality. 
This non-trivial prediction can potentially be probed in polarized neutron scattering experiments.

In passing we note that other more exotic spin liquid phases with (possibly gapless) fermionic spinons 
could potentially exist as stable phases\cite{wen91,bfn,z2long,stableu1} 
but the transitions to the Neel state have not been studied. 
The corresponding critical theories are also likely to be more exotic - we will therefore defer consideration
of such spin liquid phases and the corresponding critical points for the future and focus here on the 
simpler case with gapped bosonic spinons.

A controlled calculation in which both the magnetically ordered state and such a 
gapped spin liquid state appear is provided by considering 
a large-$N$ generalization of the $S=1/2$ 
Heisenberg model \cite{chungln} on the anisotropic triangular lattice. 
Specifically, the large-$N$ limit of a bosonic Sp($N$) 
Heisenberg model \cite{readsachdev,sachdevfru} was used to
obtain the mean field phase diagram as a function
of the ratio of intra-plane anisotropic couplings, $J'/J$, and 
the strength of quantum fluctuations, $\kappa=``2S"$, 
where $\kappa$ plays the same role as $2S$ (= value of spin at each site) in the SU(2) limit.
It was found that, for a range of $J'/J $, the semiclassical limit
(large $\kappa$) leads to a spiral magnetic order.
As the quantum fluctuations become 
stronger (small $\kappa$), a spin liquid state ($Z_2$ spin liquid) 
with gapped deconfined bosonic spinons emerges.

The possibility that \material is proximate to the quantum critical point between the incommensurate spiral 
and spin liquid states will affect the behavior at intermediate energy/length scales. 
As mentioned above Chubukov \etal\ \cite{chubukov} showed quite generally that the transition was in the 
universality class of the $O(4)$ fixed point in  
three euclidean dimensions. This is much higher symmetry than in the 
original microscopic model. This enlarged $O(4)$ symmetry acts naturally on spinon degrees 
of freedom which thus emerge as the useful variables already at the transition to the 
spin liquid. The theory predicts a large anomalous exponent $\bar{\eta}$
for the spin-spin correlation function since the spins  
are composite operators in terms of spinon operators. 
Such composite operators usually have large anomalous 
dimensions. A qualitative physical picture is simply that the spin-$1$ magnons decay rapidly 
into the spinons thereby leading to broad lineshapes in the inelastic 
neutron scattering. However right at the critical point the spinons are not free particles. 
From the large-$N$ calculation extrapolated to the physical case of $SU(2)$ spins, 
the anomalous exponent may be estimated to be 1.54\cite{chubukov}. 
A more accurate value of $\bar{\eta}$ can be obtained directly by 
classical Monte Carlo simulations of the $O(4)$ nonlinear 
sigma model in three dimensions. In the present work and 
in Ref.~\onlinecite{ballesteros}, it is found that $\bar{\eta}=1.373$. 
It is worth noting that the neutron experiments\cite{coldeaneu} in \material were fit 
to functional forms that also suggest large anomalous dimension $\bar{\eta}_{\text{E}}$
in the range 0.7 to 1. There is however reason to question these measurements 
of the actual numerical value of $\bar{\eta}$ as we discuss in Sec~\ref{sec:experimental}. Nevertheless the large anomalous 
dimension predicted by the theory is qualitatively consistent with the broad lineshapes seen in experiment.

Thus it is certainly desirable to have other qualitatively distinct 
predictions of the critical theory. It is our hope that the enlarged $O(4)$ symmetry 
and the consequent enhanced 
vector spin chirality correlations will provide such a 
sharp test. We also outline some other less dramatic consequences of the extra $O(4)$ symmetry 
that too may be useful. 

From a theoretical point of view our considerations are rather similar to those in a recent study\cite{su4} of 
``algebraic spin liquid'' phases where too the low energy theory is characterized by non-trivial enlarged 
symmetry as compared to the microscopic model. There as in the present problem this enlarged symmetry 
acts naturally on ``spinon'' degrees of freedom and leads to non-trivial relationships between the fluctuations of 
rather different competing orders. Perhaps such phenomena are common in correlated systems.

The rest of this paper is organized as follows. 
In Sec.~\ref{sec:model}, we introduce the model that we use to 
describe Cs$_2$CuCl$_4$. We briefly review the 
$O(4)$ invariant critical theory in Sec.~\ref{sec:critical} and 
discuss the anomalous exponent in Sec.~\ref{sec:anomalous}.
In Sec~\ref{sec:chirality}, we study the vector spin-chirality and 
scalar-chirality correlation functions. 
In Sec~\ref{sec:experimental}, we discuss possible experiments 
to test our predictions and the critical theory. Finally, we summarize 
our results and conclude in Sec~\ref{sec:conclusion}.

\section{Model}
\label{sec:model}

\begin{figure}[ht]
{
\centerline{\includegraphics[angle=0, width=2.0in]{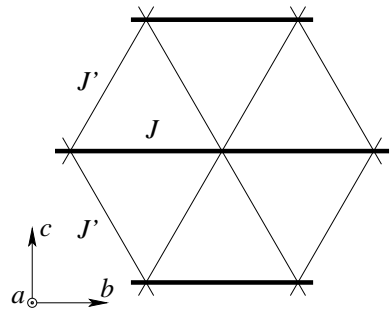}}
\caption{
Schematic representation of the triangular lattice in the $b$-$c$ plane. 
The $a$ direction is perpendicular to the plane. 
There are two coupling constants; $J$ along the horizontal chains 
and $J'$ along the zig-zag bonds.
}
\label{fig:tri}
}
\end{figure}

The magnetic ions Cu$^{2+}$ in \material carry $S=1/2$ spin
moments that reside on a stack of triangular layers \cite{coldeaneu}. 
The intralayer antiferromagnetic interaction (the interaction in the 
$b$-$c$ plane; see Fig.1) is anisotropic with two coupling constants: 
$J \approx 0.375 meV$ along the $b$ direction and $J' \approx J/3$ along the zig-zag 
bonds \cite{coldeaneu} (see Fig.~\ref{fig:tri}).
The interlayer interaction (the interaction along
the $a$ direction) is weak\cite{coldeaneu} with a coupling 
constant $J''=0.045J\ll J$. 
Therefore the system is quasi two-dimensional and can be
modeled by a two-dimensional frustrated Heisenberg Hamiltonian 
on the triangular lattice. The main part of the Hamiltonian reads
\beq
  H = J\sum_{\langle i,j \rangle} \vecb{S}_{i}  \cdot \vecb{S}_j
      + J'\sum_{\langle \langle i,j \rangle \rangle} \vecb{S}_{i} \cdot \vecb{S}_j.
\label{eq:ham}
\eeq
Here $\vecb{S}_i$ are spin-$1/2$ operators at the sites $i$ of a two dimensional triangular lattice.
The first sum runs over bonds in the $b$ direction and the 
second sum runs over the zig-zag bonds. There is also a weak 
Dzyaloshinskii-Moriya interaction \cite{coldeaham}. 
We will neglect this term when we consider the critical theory
and discuss later its role in experiments.

The weak interlayer interaction stabilizes magnetic long range order below
$T=0.62$K \cite{coldeaold}. The order is incommensurate because 
of the frustrated anisotropic interactions in the triangular planes 
and occurs at an incommensurate wave vector
$\vecb{Q}=(0.5+\epsilon_{0})\vecb{b}^*$ with
$\epsilon_{0}=0.030(2)$ and $\vecb{b}^*=(2\pi/b,0,0)$.  
The corresponding classical ordering wavevector is 
$\vecb{Q}=(0.5+\epsilon_{c})\vecb{b}^*$ with 
$\epsilon_{c}=(1/\pi)\arcsin(J'/2J)=0.053$. The substantial difference
is due to large quantum fluctuations that renormalize the ordering 
wave vector.

\subsection{Order parameter}
The spiral ordering pattern may be represented using two orthonormal unit
vectors $\vecb{n}_1$ and $\vecb{n}_2$ through:
\beq
  \langle \vecb{S}(\vecb{r}_i) \rangle \propto
    \vecb{n}_1 \cos(\vecb{Q}\cdot\vecb{r}_i)
    + \vecb{n}_2 \sin(\vecb{Q}\cdot\vecb{r}_i),
\label{eq:ordpatt}
\eeq
where $\vecb{n}_1$ and $\vecb{n}_2$ satisfy
\beq
  \vecb{n}_1\cdot\vecb{n}_2=0, \quad \vecb{n}_1^2=\vecb{n}_2^2=1.
\label{eq:constr}
\eeq

The $\vecb{n}_{1,2}$ are clearly vectors under global spin rotation, and together they define the 
order parameter for the spiral state. In addition to global spin rotations the 
spin Hamiltonian Eq.~\ref{eq:ham} is invariant under various lattice symmetries and
time reversal. It is useful and straight forward to work out the  transformation properties 
of $\vecb{n}_{1,2}$ under these symmetries. We find
\bea
  T_{\vecb{\hat{l}}}  &:& (\vecb{n}_1+i\vecb{n}_2)\rightarrow
    (\vecb{n}_1+i\vecb{n}_2)e^{i\vecb{Q} \cdot \vecb{\hat{l}}},  \nonumber \\
  R_c,I &:& (\vecb{n}_1+i\vecb{n}_2)\rightarrow(\vecb{n}_1-i\vecb{n}_2),
     \nonumber \\
  T     &:& (\vecb{n}_1+i\vecb{n}_2)\rightarrow -(\vecb{n}_1-i\vecb{n}_2).
\label{eq:nlatsym}    
\eea
Here $T_\vecb{\hat{l}}$ is a unit translation along a lattice vector $\vecb{l}$ direction, $R_c$ is reflection about the $c$ axis, 
$I$ is lattice inversion, and $T$ is time reversal. Other symmetry operations may be obtained as combinations 
of the above.

\section{Theory of quantum transition}
\label{sec:critical}


In this section we briefly review the theory of Ref.~\onlinecite{chubukov} for the quantum transition between the Neel 
and spin liquid states. We will also use this as an opportunity to clarify several possible confusions with other 
superficially similar results in the literature.

The set $\vecb{n}_1, \vecb{n}_2$ defines the order parameter for the spiral state. To study phase transitions 
out of the spiral 
state it is necessary to allow fluctuations where $\vecb{n}_{1,2}$ vary slowly as 
a function of space and time. The symmetry properties in Eq.~\ref{eq:nlatsym} actually imply 
that the resulting continuum theory has an extra $U(1)$ symmetry over and above that of spin rotations. This may be seen 
by considering translations along the $\vecb{b}$ direction. We have
\begin{equation}
(\vecb{n}_1+i\vecb{n}_2)\rightarrow(\vecb{n}_1+i\vecb{n}_2)
       e^{imQ_x}
\end{equation}
for a translation by $m$ lattice sites. With incommensurate $Q_x$, this effectively translates into a 
full $U(1)$ symmetry that rotates between the $\vecb{n}_1$ and $\vecb{n}_2$ fields.

The order parameter manifold defined by $\vecb{n}_{1,2}$ allows 
for topological vortex defects with a discrete $Z_2$ character\cite{z2}. These $Z_2$ vortices 
have energy logarithmic in the system size
in the ordered state due to the long distance distortion of the order parameter. 
Now consider disordering the spiral order while keeping the core energy of 
these $Z_2$ vortices finite. Ref. \onlinecite{chubukov} argues that the resulting state is a fractionalized spin liquid 
with bosonic spin-$1/2$ spinons. The $Z_2$ vortices survive into the spin liquid phase but now only cost 
finite (not divergent) energy. 

A theory for this transition is obtained by writing
\begin{equation}
\vecb{n}_1 + i\vecb{n}_2 = \epsilon_{\alpha \gamma} z_{\gamma} \boldsymbol{\sigma}_{\alpha \beta} z_{\beta},
\label{eq:param}
\end{equation}
where $\epsilon$ is the antisymmetric tensor, $\sigma^a$ are the Pauli
matrices, and $z_{\alpha} = (z_\uparrow, z_\downarrow)$ a two component complex unit vector satisfying
\begin{equation}
z^{\dagger}z = 1.
\end{equation}
It is easy to check that the parametrization (\ref{eq:param}) satisfies
the constraints given by Eq.~\ref{eq:constr}.
The $z_{\alpha}$ transform as spinors under the $SU(2)$ spin rotation and describe
spin-$1/2$ spinons. This representation clearly has a $Z_2$ gauge redundancy associated with changing the 
sign of $z$ at any point in space-time. 
Thus a reformulation in terms of $z$ may be fruitfully viewed as a theory of the $z$ fields
coupled to a $Z_2$ gauge field.  The corresponding $Z_2$ gauge flux is associated with the $Z_2$ 
vortices discussed above. These vortices stay gapped at the transition between the Neel and spin liquid states
and may hence be ignored for a low energy description of the critical point. 
Thus we may obtain the critical behavior by focusing only on the spinons and ignoring their coupling to 
the $Z_2$ gauge field. Detailed arguments show that the critical theory is in fact in the $O(4)$ universality class
in $D = 2+1$ dimensions where there is full rotational symmetry between the four real numbers described by 
$(z_{\uparrow}, z_{\downarrow})$. In effect the extra $U(1)$ symmetry of rotation between 
$\vecb{n}_1$ and $\vecb{n}_2$ has been enlarged to $SU(2)$. Combined with the $SU(2)$ spin rotations
the full symmetry is 
$SU(2) \times SU(2) \sim O(4)$. Thus the critical properties may be computed using the Euclidean action
\beq
  {\cal S }=\int \text{d}^2 x \, \text{d}\tau \sum_{\mu=x,\tau}
    \frac{1}{g}
      \partial_{\mu} z_{\alpha}^* \partial_{\mu} z_{\alpha}.
\eeq

The relationship between the $z$ and $\vecb{n}_{1,2}$ fields may also be 
expressed in a different way that will be fruitful later. 
Let us introduce an $SU(2)$ matrix $U$ built out of $z$:
\beq
  U=\left(
  \matrix{
    z_{\uparrow} & z_{\downarrow}^* \cr
    z_{\downarrow} & -z_{\uparrow}^* \cr
  }
  \right).
\eeq
Then $U$ satisfies
\begin{equation}
U^{\dagger} \sigma^a U = R^{ab} \sigma^b,
\end{equation}
where $R$ is a $3 \times 3$ rotation matrix. Clearly
\begin{equation}
R^{ab} = \frac{1}{2}tr\left(U^\dagger \sigma^a U \sigma^b \right).
\end{equation}
It is readily checked that $\vecb{n}_1$ and $\vecb{n}_2$ are the first and second column 
respectively of this rotation matrix. The third column represents the unit vector 
\begin{equation}
\vecb{n}_3 = \vecb{n}_1 \times \vecb{n}_2
\end{equation}
which is orthogonal to both $\vecb{n}_1$ and $\vecb{n}_2$. Symmetry under physical spin rotations correspond to 
{\em left} multiplication of $U$ by an $SU(2)$ spin rotation matrix:
\begin{equation}
U \rightarrow VU
\end{equation}
with $V \in SU(2)$. The enlarged $O(4)$ symmetry implies that {\em right} multiplication 
\begin{equation}
 U \rightarrow UV 
\end{equation}
is also a symmetry of the critical fixed point.

Several comments are in order on these results. Prior to Ref.~\onlinecite{chubukov}, analysis of a continuum 
non-linear sigma model appropriate for non-collinear magnets had led to the suggestion of enlarged 
$O(4)$ symmetry in $2+ \epsilon$ space-time dimensions\cite{azaria}. However numerical calculations on stacked triangular
lattices near their finite temperature ordering transition failed to observe the predicted $O(4)$ 
universality class\cite{kwmrrev}. 
Rather the evidence supports a transition in the ``chiral'' universality class of Ref.~\onlinecite{kwmr,kwmrrev}. 
How then are we to reconcile this with our claims about the quantum Neel - spin liquid transition being in the $O(4)$ 
universality class? The answer lies in the nature of the paramagnetic phase. 
In the classical stacked triangular lattice 
the natural paramagnetic phase is the trivial one that is smoothly connected to the high temperature limit. 
The transition to this phase from the ordered state is obtained by proliferating the $Z_2$ vortices (which are 
line defects in the three dimensional classical model). In principle an exotic paramagnetic phase is also possible 
which would be the classical three dimensional analog of the spin liquid: this requires destroying the magnetic order 
without proliferating the $Z_2$ vortices. This paramagnet will be topologically ordered and will be separated from 
the trivial very high temperature paramagnet by a phase transition. In this classical context, 
the arguments of Ref.~\onlinecite{chubukov} apply to the transition between the Neel state and 
this topologically ordered 
paramagnet which will indeed be in the $O(4)$ universality class. However neither this non-trivial paramagnet 
nor the corresponding transition were apparently accessed in the numerical calculations\cite{kwmrrev}. 
Equally it is hard 
to decide which of the two possible paramagnetic phases were accessed in the $2 + \epsilon$ calculations of 
Ref. \onlinecite{azaria}. 
The distinction between the two phases is topological and hence sensitive to spatial dimension. This is hard to 
disentangle in the $\epsilon$ expansion. 

Turning to the quantum problem at hand it differs from the classical stacked magnet in an important way. There are 
extra Berry phases that are sensitive to the microscopic spin at each lattice site 
({\em i.e.} spin-$1/2$ or spin-$1$, etc). A close and familiar analogy is from the theory of 
collinear quantum antiferromagnets in two dimensions where such Berry phases spoil any general direct mapping to 
classical collinear magnets in one higher dimension. In that case in the semiclassical limit the Berry phases 
are associated entirely with singular topological configurations known as hedgehogs. 
A similar result also holds for the non-collinear quantum magnets of interest in this paper. The Berry phases 
are associated entirely with the topological $Z_2$ vortex configurations. As these $Z_2$ vortices 
are gapped across the 
transition to the spin liquid the Berry phases play no role in the low energy universal critical physics. Indeed the 
$O(4)$ universality class will describe the Neel - spin liquid transition for all spin-$S$ magnets 
regardless of the value of $S$. On the other hand the phase obtained when the Neel state is 
disordered by proliferating the 
$Z_2$ vortices will be strongly influenced by the Berry phases. For spin-$1/2$ it is expected that the Berry phases will 
lead to broken lattice symmetries in the resulting paramagnet\cite{sen}. The nature of the transition 
between such a VBS ordered 
paramagnet and the Neel state on the triangular lattice is not presently understood. Preliminary analysis suggests 
that an interesting ``Landau-forbidden" second order transition may be possible. However in the present paper we 
restrict ourselves to studying the transition to the spin liquid.

\section{Consequences of critical theory}

\subsection{Spin correlations}
\label{sec:anomalous}

We begin by considering the spin-spin correlation function at the ordering wavevector
which is readily accessed in neutron scattering experiments. 
This can be expressed in terms of the
correlation functions of the vectors $\vecb{n}_1$ and $\vecb{n}_2$
\beq
  \langle \vecb{S}(\vecb{r},\tau) \cdot \vecb{S}(0,0) \rangle =
    \langle \vecb{n}_1(\vecb{r},\tau) \cdot \vecb{n}_1(0,0) \rangle
    \cos(\vecb{Q} \cdot \vecb{r}).
\label{eq:corr}
\eeq
We note that due to the symmetry of rotations between $\vecb{n}_1$ and $\vecb{n}_2$, 
they will both have the same correlations. 
Close to the critical point, the correlation function has the 
usual power-law behavior
\beq
  \langle \vecb{S}(-\vecb{q},\omega) \cdot \vecb{S}(\vecb{q},\omega)
    \rangle \sim \frac{1}{(\omega^2-k^2)^{1-{\bar{\eta}}/2}},
\eeq
where ${\bf k}={\bf q}-{\bf Q}$.
Since $\vecb{n}_1$ and $\vecb{n}_2$ (and $\vecb{S}$) are composite
operators in terms of the $z_{\alpha}$ fields, one can expect that the
spin-spin correlation function has a large anomalous exponent $\bar{\eta}$.
Indeed, the large-$N$ calculation of $\bar{\eta}$ for an $O(2N)$ theory
gives \cite{chubukov}
\beq
  \bar{\eta}=1+\frac{32}{3\pi^2N}.
\eeq
For the physical case $N=2$, we have $\bar{\eta}\approx1.54$. 
This is a large number compared to anomalous exponents of 
non-composite operators.

To improve the large-$N$ estimate of the anomalous exponent, one can alternately
perform Monte Carlo simulations of the classical $O(4)$ nonlinear sigma model
in three dimensions since this model is in the same universality class as the
transition of interest. Ballesteros \etal\ \cite{ballesteros} measured the anomalous
exponent $\eta_{\text{T}}$ of the tensorial magnetization that is defined as
\bea
  M_{\text{T}}&=& \left \langle \sqrt{\text{Tr}{\cal M}^2} \right \rangle, \nonumber \\
  {\cal M}_{\alpha\beta} &=& \sum_i (\phi_{i\alpha} \phi_{i\beta}
    -\frac{1}{4} \delta_{\alpha\beta}), \nonumber
\eea
where $\phi_{i\alpha}$ are four components of the $O(4)$ unity vector
at the site $i$ in three-dimensions.
They found $\eta_{\text{T}}=1.375(5)$. This result is directly related
to our case because of the following argument. Using Eq.~\ref{eq:param}
and writing $z_{\uparrow}=(\phi_1+i\phi_2), z_{\downarrow}=(\phi_3+i\phi_4)$,
where $\phi_{\alpha}$ are now real fields, we have
\bea
  \vecb{n}_1&=&\left(
    \matrix{
      \phi_2^2+\phi_3^2-\phi_1^2-\phi_4^2 \cr
      2(\phi_1\phi_2+\phi_3\phi_4) \cr
      2(\phi_1\phi_3-\phi_2\phi_4) \cr
    }
  \right) \label{eq:n1} \\
  \vecb{n}_2&=&\left(
    \matrix{
      2(\phi_3\phi_4-\phi_1\phi_2) \cr
      \phi_2^2+\phi_4^2-\phi_1^2-\phi_3^2 \cr
      2(\phi_1\phi_4+\phi_2\phi_3) \cr
    }
  \right).
\label{eq:n2}
\eea
We can see that the tensorial magnetization is related to $\vecb{n}_1$
and $\vecb{n}_2$ by symmetry and therefore the anomalous exponent of the
tensorial magnetization is equal to the anomalous exponent of the
vectors $\vecb{n}_1$ or $\vecb{n}_2$.
Then, from Eq.~\ref{eq:corr}, it is the same as the anomalous exponent 
of the correlation function at wavevector $\vecb{Q}$.

We have confirmed this conclusion by computing
the anomalous exponent of $\vecb{n}_1$ directly. We simulate the $O(4)$
nonlinear sigma model in three dimensions on the simple cubic lattice using
a cluster algorithm \cite{wolff}. We measure the anomalous exponent
$\bar{\eta}$ in the following way. We calculate the total $z$ component
$n_1^z=2(\phi_1\phi_3-\phi_2\phi_4)$ of the vector $\vecb{n}_1$
\beq
  M=\sum_i n_{1i}^z,
\eeq
where the sum runs over all lattice sites. In our simulations, we measure
the ``magnetization'' and ``susceptibility``
\bea
  \bar{m}&=&\frac{1}{L^3} \langle M \rangle,\nonumber \\
  \bar{\chi}&=&\langle M^2 \rangle,\nonumber
\eea
where $L$ is the system size. We perform our simulations only at the
critical temperature found in Ref.~\onlinecite{ballesteros} with very high
accuracy, $\beta_c=1/T_c=0.935861(8)$. To find the critical exponents, we
use finite size scaling analysis. Close to the critical point, observables
satisfy the general scaling form
\beq
        {\cal O}=L^{\rho/\nu} F_{\cal O}(tL^{1/\nu},L^{-\omega}),
\eeq
where ${\cal O}$ can be $\bar{m}$ or $\bar{\chi}$, $\rho$ is
the scaling exponent corresponding to the operator ${\cal O}$
($\rho=-\bar{\beta}=-\nu(1+\bar{\eta})/2$ for $\bar{m}$ and
$\rho=\bar{\gamma}=\nu(2-\bar{\eta})$ for $\bar{\chi}$), $F_{\cal O}$ is a
universal function, $t=|T-T_c|$, and $\omega$ is a universal exponent
related to the leading irrelevant operator. Exactly at the critical point $t=0$,
one can write
\beq
        {\cal O}=L^{\rho/\nu} (a+cL^{-\omega}).
\eeq
We assume that the correction to scaling is negligible for large lattice
sizes and fit to a simplified scaling form
\beq
        {\cal O}=aL^{\rho/\nu}.
\eeq
We have performed simulations for large enough lattices (up to
$L_{\text{max}}=96$). Fitting from $L_{\text{min}}=20$ to $L_{\text{max}}$,
we obtain $\bar{\eta}=1.373(3)$.

The large value of $\bar{\eta}$ is a reflection of the emergence of spinons as useful 
degrees of freedom at the critical point.

\subsection{Other competing orders}
\label{sec:chirality}

The extra $O(4)$ symmetry actually has more striking consequences for the critical properties. 
It implies that operators other than the natural magnetic order parameter will have enhanced 
power law correlators. Consider the magnetic order parameters $\vecb{n}_{1,2}$. As discussed in 
Section \ref{sec:critical}, they may be regarded as the first and second columns of a
rotation matrix $R$. The third column of the rotation matrix is 
 $\vecb{n}_3 =\vecb{n}_1 \times \vecb{n}_2$. The enlarged $O(4)$ symmetry at the critical point implies 
that both left and right multiplications by an orthogonal matrix of this rotation matrix $R$ 
are symmetries of the critical theory. 
Left multiplication is just physical spin rotation. However symmetry under right multiplication implies that 
$\vecb{n}_{1,2,3}$ will all have the same correlations. In particular $\vecb{n}_3$ will have the same slow power law
decay as $\vecb{n}_{1,2}$ calculated in the previous subsection. 
We can
express the vector $\vecb{n}_3$ in terms of the spin operator as follows.
Consider $\vecb{S}_{\vecb{r}+\vecb{\hat{l}}} \times \vecb{S}_{\vecb{r}}$, where
$\vecb{\hat{l}}$ is a unit lattice vector. Using (\ref{eq:ordpatt}) and expanding
the first spin operator around $\vecb{r}$, we have
\bea
  \vecb{S}_{\vecb{r}+\vecb{\hat{l}}} \times \vecb{S}_{\vecb{r}} &\sim&
    \vecb{n}_3 \sin(\vecb{Q}\cdot\hat{\vecb{l}}) \nonumber \\
    &+& \text{derivative terms}.
\label{eq:sts}
\eea
We can ignore the derivatives terms since their contribution to the correlation
function is negligible. Consider the chirality vector
\beq
  \vecb{C}_{\vecb{r}}=
    \sum_{\vecb{r}'} \vecb{S}_{\vecb{r}} \times \vecb{S}_{\vecb{r}'},
\eeq
where the oriented sum is over an elementary plaquette. Using
Eq.~(\ref{eq:sts}), it is easy to show that
\beq
  \vecb{C}=\left[\sin Q_x-2\sin \frac{Q_x}{2} \cos \frac{\sqrt{3}Q_y}{2}\right]
    \vecb{n}_3 \propto \vecb{n}_3.
\eeq
Thus the vector spin-chirality correlation function
\bea
  \langle \vecb{C}(\vecb{r},\tau) \cdot \vecb{C}(\vecb{0},0)\rangle &\propto&
    \langle \vecb{n}_3(\vecb{r},\tau) \cdot \vecb{n}_3(\vecb{0},0)\rangle \nonumber \\
    &=& \langle \vecb{n}_1(\vecb{r},\tau) \cdot \vecb{n}_1(\vecb{0},0)\rangle.
\eea
have the same power-law behavior at zero wavevector as those of spins
at the wavevector $\vecb{Q}$. This highly non-trivial statement possibly provides a 
way to test the theory of Ref.~\onlinecite{chubukov} in 
experiments and in model numerical calculations.

\subsection{Conserved quantities}
Some less dramatic effects of the enlarged $O(4)$ symmetry are also worth noting. 
The $O(4)$ group has six generators
and these will all be conserved at the critical fixed point. Three of these are just the conserved total spin
$\vecb{S}_{tot} = \sum _{i} \vecb{S}_i$ - they are the generators of left rotations of $U$. The remaining three 
are conserved only in the low energy critical theory (though not in the original microscopic model). They correspond to 
right rotations of $U$. We will denote these $K_a$, $ a = 1,2,3$. 
The $K_a$ transform as vectors under the group 
$SU(2)_R$ of right rotations of $U$. They generate rotations of $\vecb{n}_a$ amongst one another so that
\begin{equation}
\left[\vecb{n}_a, K_b \right] = i \epsilon_{abc} \vecb{n}_c.
\label{eq:comm}
\end{equation}

As is well-known such conserved quantities have scaling dimension $d_{\rm scale} = 2$ with no anomalous 
dimension\cite{footnote}.
Thus at the critical point at zero temperature
\begin{equation}
\langle K_a ({\bf r}, \tau)K_b({\bf 0},0) \rangle \sim \frac{\delta_{ab}}{(|{\bf r}|^2 + \tau^2)^2},
\end{equation}
where ${\bf r}$ is the spatial coordinate and $\tau$ is the (imaginary) time coordinate. A similar result also holds for 
the conserved 
total spin. Away from the critical point at finite temperature in the quantum critical region, the $K_a$ 
will continue to be approximately conserved. The exact conservation will be spoiled by irrelevant operators that 
break the $O(4)$ symmetry down to $SU(2) \times lattice ~space ~group$. Thus at $T > 0$ in the quantum critical region
we expect that the $K_a$ will diffuse upto a long length/time scale which will be determined by 
$T$ and the distance to the critical point. It is possible that the presence of such extra nearly diffusive modes 
can also be looked for in experiments. 

But what do the $K_a$ correspond to in terms of the underlying spins? A useful guess for the answer 
is provided by examining the transformation properties of $K_a$ under the microscopic symmetries of the 
original lattice system. It is reasonable that any lattice operator that has the same transformation
properties will have some overlap in the continuum theory with the $K_a$. With the further assumption that
there are no other operators in the continuum theory with smaller scaling dimension that also have the same 
transformation, the $K_a$ will give the dominant correlations of such lattice operators. 
Similar considerations were also used in Ref.~\onlinecite{su4}.

The transformation properties of $K_a$ under all physical symmetries is fixed by the
commutation relation Eq.~\ref{eq:comm}.
The transformation properties of $K_a$ are related to the transformation
properties of the matrix $R$. First, we list the latter ones.
\bea
  SU(2)_{\text{spin}} &:& R\rightarrow OR, \nonumber \\
  T_{\vecb{\hat{l}}} &:&
    R \rightarrow R
    \left(
    \matrix{
      \cos(\vecb{Q}\cdot\vecb{\hat{l}}) &
         \sin(\vecb{Q}\cdot\vecb{\hat{l}}) & 0 \cr
      -\sin(\vecb{Q}\cdot\vecb{\hat{l}}) &
         \cos(\vecb{Q}\cdot\vecb{\hat{l}}) & 0 \cr
      0 & 0 & 1
    }
    \right), \nonumber \\
  R_c,I &:&
    R \rightarrow R
    \left(
    \matrix{
       1 &  0 &  0 \cr
       0 & -1 &  0 \cr
       0 &  0 & -1
    }
    \right), \nonumber \\
  T &:&
    R \rightarrow R
    \left(
    \matrix{
      -1 &  0 &  0 \cr
       0 & -1 &  0 \cr
       0 &  0 &  1
    }
    \right), \nonumber
\eea
where $O$ is an $O(3)$ matrix and $T_{\vecb{\hat{l}}}$ is the lattice
translation along the lattice vector $\vecb{\hat{l}}$. We can use these
transformations and the condition that $K_a$ are vectors under $SU(2)_R$ to
obtain the transformations of $K_a$
\bea
  SU(2)_{\text{spin}}  &:& (K_1,K_2,K_3)\rightarrow (K_1,K_2,K_3), \nonumber \\
  T_{\vecb{\hat{l}}} &:& (K_1,K_2,K_3)\rightarrow (K_1,K_2,K_3)M, \nonumber \\
  R_c,I &:& (K_1,K_2,K_3)\rightarrow (K_1,-K_2,-K_3), \nonumber \\
  T     &:& (K_1,K_2,K_3)\rightarrow (K_1,K_2,K_3), \nonumber
\eea
where
\beq
  M=\left(\matrix{
      \cos(\vecb{Q}\cdot\vecb{\hat{l}})
        & -\sin(\vecb{Q}\cdot\vecb{\hat{l}}) & 0 \cr
      \sin(\vecb{Q}\cdot\vecb{\hat{l}})
        & \cos(\vecb{Q}\cdot\vecb{\hat{l}}) & 0 \cr
      0 & 0 & 1
    }\right).
\eeq

We can construct local spin operators that have the same symmetry properties
as $K_a$. We find that
\bea
  K_1(\vecb{r}) &\sim& \cos (\vecb{Q}\cdot\vecb{r}) \vecb{S}(\vecb{r}) \cdot
    [\vecb{S}(\vecb{r}-\vecb{\hat{x}}) + \vecb{S}(\vecb{r}+\vecb{\hat{x}})], \\
  K_2(\vecb{r}) &\sim& \sin (\vecb{Q}\cdot\vecb{r}) \vecb{S}(\vecb{r}) \cdot
    [\vecb{S}(\vecb{r}-\vecb{\hat{x}}) + \vecb{S}(\vecb{r}+\vecb{\hat{x}})], \\
  K_3(\vecb{r}) &\sim& \vecb{S}(\vecb{r}-\vecb{\hat{x}}) \cdot
    [\vecb{S}(\vecb{r}) \times \vecb{S}(\vecb{r}+\vecb{\hat{x}})].
\eea
Thus $K_1+iK_2$ transforms identically with the Fourier transform of the bond energy at wavevector
$\vecb{Q}$ while  $K_3$ can be identified with the scalar spin-chirality
along the chains. 

\section{Experimental implications}
\label{sec:experimental}

\subsection{Spin correlations}
 
Quite generally in the quantum critical region the spin correlations at the ordering wavevector 
must satisfy scaling - for instance as a function of $\omega/T$.  \material orders at low temperature. 
Consequently  
data at the lowest temperatures
and frequencies must be excluded from the scaling analysis. Furthermore it is precisely the 
very-low-$T$/low-$\omega$ range that will also be most sensitive to the presence of 
weak spin anisotropies (such as Dzyaloshinski-Moriya interactions). However the data at intermediate $T$
and $\omega$ will be less affected by such spin anisotropies or by the low energy long range order.

The value of $\bar{\eta}$ found from Monte Carlo simulations 
can be compared to $\bar{\eta}_E$ estimated in experiments 
using a fitting procedure \cite{coldeaneu}. The experimental
values fall in a range between 0.7 and 1. Thus the Monte-Carlo 
result is still larger than the experimentally estimated value.
It is important to note the following caveat on this comparison.
The existing measurements of $\bar{\eta}_E$ were performed in inelastic neutron scattering experiments 
where {\em both the wavenumber and frequency transfer to the sample were simultaneously varied}. 
In particular typical data sets (such as for instance scan G of Ref.~\onlinecite{coldeaneu}) involve 
increasing the frequency transfer while the momentum transfer varies from points in the two-dimensional 
Brillouin zone far away from the ordering wavevector $\vecb{Q}$ to points close to $\vecb{Q}$. 
In the context of the ideas explored in this paper the frequency dependence at fixed $\vecb{q}$ far away from 
$\vecb{Q}$ will not be dominated by the singular long distance critical fluctuations and will be highly 
non-universal. In contrast the frequency dependence for $\vecb{q}$ close to $\vecb{Q}$ will be 
determined by the long distance critical fluctuations and will be universal. It is these latter fluctuations 
that are described by the calculated exponent $\bar{\eta}$. Thus the existing measurements of $\bar{\eta}_E$
are quite possibly severely contaminated by non-universal short distance effects. Hence the lack of 
quantitative agreement between $\bar{\eta}$ and $\bar{\eta}_E$ is not surprising. However 
the qualitative observation of broad magnon linewidths is consistent with the large $\bar{\eta}$ 
obtained in the theory. Future experiments will hopefully directly probe the fluctuations near the 
ordering wavevector thereby allowing for quantitative comparison. 

As noted already in Ref.~\onlinecite{chubukov}, NMR experiments may be a useful way to directly 
measure $\bar{\eta}$. For the nuclear relaxation rate we have 
\begin{equation}
\frac{1}{T_1} \sim T^{\bar{\eta}}.
\end{equation}
Again this behavior will only obtain at intermediate temperatures.

\subsection{Detection of vector spin-chirality fluctuations}
A dramatic consequence of the critical theory studied in this paper is the enhanced vector spin-chirality
correlations discussed in Section \ref{sec:chirality}. How can they be detected in experiments? In this section we 
show how this may be done through 
polarized neutron scattering. Our proposal builds on and generalizes the pioneering ideas of Maleyev\cite{Maleyev}
in the context of classical non-collinear magnets. 

The part of the neutron scattering rate $R_P$ that depends on the incoming neutron polarization $\vecb{P}_i$ is 
given by
\begin{equation}
R_P(\vecb{k}_i, \vecb{P}_i \rightarrow \vecb{k}_f) \sim \left(\vecb{P}_i \cdot \hat{q} \right) 
\left(\hat{q} \cdot \vecb{b} \right),
\end{equation}
where $\vecb{q} = \vecb{k}_f - \vecb{k}_i$ is the three-momentum transferred to the sample, and $\hat{q}$ is 
the corresponding unit vector. The vector $\vecb{b}$ is given by 
\begin{equation}
b_{\alpha}  =  \frac{\epsilon_{\alpha\beta\gamma}}{2i}S_{\beta \gamma},
\end{equation}
where $S_{\beta\gamma}$ is the spin structure factor. Thus the polarization dependent part 
probes the {\em antisymmetric} part of the structure factor.
By the usual arguments it is clear that $b_{\alpha}$ can be obtained from a calculation of the imaginary time 
Green function (here $0 < \tau < \beta=1/T$)
\begin{eqnarray}
\vecb{g}(\vecb{r},\tau) &  = & \langle \vecb{S}(\vecb{r},\tau) \times \vecb{S}(\vecb{0},0) \rangle, \\
\vecb{g}(\vecb{q}, i\omega_n) & = & \int_{r\tau} e^{i\vecb{q} \cdot \vecb{r} 
-i\omega_n \tau} \vecb{g}(\vecb{r},\tau).
\end{eqnarray}
Specifically we have 
\begin{equation}
\vecb{b}(\vecb{q},\omega) = \frac{-1}{1-e^{-\beta\omega}}\left(\frac{
\vecb{g}(\vecb{q}, i\omega_n \rightarrow \omega+i0^+) - 
\vecb{g}(\vecb{q}, \omega+i0^-)}{2} \right).
\end{equation}

\begin{figure}[ht]
{
\centerline{\includegraphics[angle=0, width=2.0in]{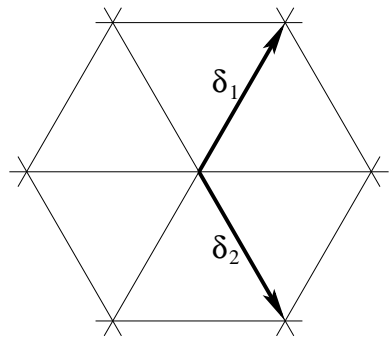}}
\caption{
Dzyaloshinski-Moriya interaction is non-zero along oriented zigzag bonds 
that are denoted by
vectors $\boldsymbol{\delta}_{1}$ and $\boldsymbol{\delta}_{2}$.
}
\label{fig:dm}
}
\end{figure}

In the context of this paper using Eq.~\ref{eq:ordpatt} we find
\begin{eqnarray}
\vecb{g}(\vecb{r}, \tau) & = & 
\langle \vecb{n}_1(\vecb{r}, \tau) \times \vecb{n}_1(\vecb{0},0) \rangle 
\cos(\vecb{Q} \cdot \vecb{r}) \nonumber \\
& + & \langle \vecb{n}_2({\bf r}, \tau) \times \vecb{n}_1(\vecb{0},0) \rangle 
\sin(\vecb{Q} \cdot \vecb{r}). \nonumber
\end{eqnarray}
Clearly such correlators are zero if there is full spin isotropy. In the specific case of \material
the presence of a weak Dzyaloshinski-Moriya(DM) interaction will lead to a non-zero result. 
It is known that the DM interaction is non-zero along oriented zigzag bonds. For a single triangular layer
\begin{equation}
H_{DM} = -\vecb{D} \cdot \sum_{\vecb{r}} 
\vecb{S}_{\vecb{r}} \times \left(\vecb{S}_{\vecb{r} + \boldsymbol{\delta}_1} + 
\vecb{S}_{\vecb{r} + \boldsymbol{\delta}_2}\right).
\end{equation}
Here $\boldsymbol{\delta}_{1,2}$ are unit vectors along the oriented zigzag bonds as shown 
in Fig.~\ref{fig:dm}. 
In addition $H_{DM}$ is also staggered between different layers. The DM vector is oriented along the $a$ axis
and has magnitude $D \approx 0.02 meV \approx 0.05 J$. In the continuum theory we may again use 
Eq.~\ref{eq:ordpatt} to write
\begin{eqnarray}
H_{DM} & \approx & -\vecb{d} \cdot \int d^2r \ \vecb{n}_1 \times \vecb{n_2}, \\
& = & -\vecb{d} \cdot \int d^2r \ \vecb{n}_3.
\end{eqnarray}
Here $\vecb{d} \propto \vecb{D}$. Let us now try to evaluate $\vecb{g}$ to linear order in $\vecb{d}$. 
We get
\begin{eqnarray}
\vecb{g}(x) & = & \int d^3x' \langle \vecb{n}_1(x) \times \vecb{n}_1(0) 
\left(\vecb{d} \cdot \vecb{n}_3(x')\right) \rangle \cos(\vecb{Q} \cdot \vecb{r}) \nonumber \\
& + & \int d^3x'\langle \left(\vecb{n}_2(x) \times \vecb{n}_1(0)  \right)
\left(\vecb{d} \cdot \vecb{n}_3(x')\right) \rangle \sin(\vecb{Q} \cdot \vecb{r}). \nonumber
\end{eqnarray}
Here $x = (\vecb{r},\tau), x' = (\vecb{r'}, \tau')$ are space-time coordinates.
The averages are evaluated in the isotropic theory. The first average vanishes due to the symmetry
$\vecb{n}_3 \rightarrow -\vecb{n}_3, \vecb{n}_1 \rightarrow \vecb{n}_1$ present in the action. 
We are therefore left with
\begin{equation}
\vecb{g}(x) = d_{\alpha}\int d^3x'\langle \vecb{n}_2(x) \times \vecb{n}_1(0)  
n_{3\alpha}(x') \rangle \sin(\vecb{Q} \cdot \vecb{r}).
\end{equation}
Thus $\vecb{g}$ is determined by the three point correlation function of $\vecb{n}_{1,2,3}$. Using 
spin rotation invariance we have
\begin{equation}
\vecb{g}(\vecb{r},\tau) = \frac{\vecb{d}}{3} \sin(\vecb{Q} \cdot \vecb{r})\int d^3x' 
\langle \vecb{n}_2(x) \times \vecb{n}_1(0) \cdot \vecb{n}_{3}(x') \rangle .
\end{equation}
Such 3-point correlators of scaling fields are severely restricted by conformal invariance\cite{ginscft}. 
As $\vecb{n}_{1,2,3}$ all have the same scaling dimension $\Delta = (1+\bar{\eta})/2$, we have (at $T = 0$)
\begin{equation}
\label{3ptcft}
\langle \vecb{n}_2(x) \times \vecb{n}_1(0) \cdot \vecb{n}_3(x') \rangle
\sim \frac{1}{x^{\Delta}x'^{\Delta}|x-x'|^{\Delta}}.
\end{equation}

We now have to integrate over $x'$ which is the coordinate of 
$\vecb{n}_3$. It is easy to see that this integral is infra-red(IR) divergent. 
This is because at large $x'$, the integrand behaves as 
$\frac{1}{x'^{2\Delta}}$.
With $2\Delta = 1 + \bar{\eta} \approx 2.37$, the integral over $x'$ will diverge in 
the infra-red. 
Formally this just means that perturbation theory in the DM interaction 
diverges at $T = 0$ and the correct answer will involve some fractional 
power of the DM vector. 
However for our purposes it is much more meaningful and simpler to go to finite temperature 
where this divergence 
will be cut-off. Before doing that it is useful 
to understand the origin of this divergence at $T = 0$ better. 

The divergence comes from large $x' \gg x$. To discuss this limit let us 
keep $x'$ fixed and bring $x$ close to $0$. Then we need 
the operator product expansion(OPE) of $\vecb{n}_2 \times \vecb{n}_1$. This is a vector in spin space 
and the leading term will
just be $\vecb{n}_3$. Scaling requires the equation
\begin{equation}
\vecb{n}_2(x) \times \vecb{n}_1(0) \sim \frac{1}{x^{\Delta}} \vecb{n}_3 (x).
\end{equation}
There is only one power of $\Delta$ in the right hand side so that both sides will scale 
identically. 
Now if we calculate the correlator with $\vecb{n}_3(x')$ we will reproduce the 
limit of the exact result Eq.~\ref{3ptcft} above when $x'$ is large. 
It is now clear that the infra-red divergence in the integral over $x'$ actually is 
nothing but the divergence 
of the uniform susceptibility of the vector spin chirality at zero temperature. 

Armed with this insight let us discuss $T > 0$. Here we will assume that we 
are interested in the scattering at an external frequency $\omega \gg T$. 
This simplifies things because then the sole effect of $T > 0$ is to cutoff 
the IR divergence without affecting the rest of the correlations. 

Using the OPE it is now clear that the answer is the Fourier transform of 
\begin{equation}
\frac{1}{x^\Delta} \chi_{ch}(T),
\end{equation}
where $\chi_{ch}(T)$ is the uniform vector spin-chirality susceptibility at temperature $T$. 
From scaling we have
\begin{equation}
\chi_{ch}(T) \sim \frac{1}{T^{3-2\Delta}}.
\end{equation}
Using this we may straightforwardly calculate the inelastic scattering rate. 
For scattering right at the ordering wavevector $\vecb{Q}$ and for $\omega \gg T$, we get 
\begin{equation}
R_P(\omega) \sim 
\frac{1}{\omega^{3- \Delta}T^{3-2\Delta}}
\end{equation}
with $2\Delta = 1+ \bar{\eta} \approx 2.37$. 

This is a definite prediction that can possibly be tested. 
In the actual experiments it will be necessary to take into account 
the staggering of the DM interaction between the different layers. Thus it is best to choose the 
$a$ component of the 3-vector $\vecb{q}$ to be $\pi$.

\section{Summary and Conclusions}
\label{sec:conclusion}

In this paper we have pursued a concrete version of the idea that Cs$_2$CuCl$_4$, 
though magnetically ordered at 
low temperature, may nevertheless 
be proximate to a spin liquid phase. Such proximity suggests 
that the intermediate length/time scale physics of \material may be governed by a 
quantum critical point between a magnetic spiral and a genuine two dimensional quantum 
spin liquid. A theory for such a quantum phase transition was obtained in Ref. \onlinecite{chubukov}
for a simple ($Z_2$) spin liquid with gapped bosonic spinons.  We showed that the spin correlations at the 
quantum critical point are characterized by a large anomalous exponent $\bar{\eta} \approx 1.37$. This is 
qualitatively consistent with the broad power law tails observed in inelastic neutron scattering in \material.
Further we showed that the enlarged $O(4)$ symmetry\cite{chubukov} at the critical fixed point 
has some remarkable consequences. The vector spin-chirality has the same slow power law decay as the 
natural magnetic order parameter. This sharp qualitative observation should be of great use in confirming
(or ruling out) the applicability of the theory of Ref. \onlinecite{chubukov} to \material. Building on  
Ref. \onlinecite{Maleyev}, 
we showed how polarized inelastic neutron scattering can be used to directly 
detect the vector spin-chirality correlations. It is our hope that future experiments will be able to use 
these results to clarify the physics behind the interesting properties of \material.

\section*{Acknowledgements}
We thank Radu Coldea, Young Lee, Diptiman Sen, and Ashvin Vishwanath for helpful discussions.
This work was supported by NSF Grant No. DMR-0308945 (TS), NSERC of Canada,
Canada Research Chair Program, and Canadian Institute for Advanced 
Research (SVI and YBK). 
TS also acknowledges  
funding from the NEC Corporation, the Alfred P. Sloan Foundation, and an award from the The Research Corporation.


\begin{thebibliography}{99}
\bibitem{footnote}
In addition there will be twelve currents associated with these 6 conserved densities 
which will also have scaling dimension 2 with no anomalous dimension. Of these 6 are just the spin currents. The 
remaining 6 are currents of the $K_a$.
\end{thebibliography}
\end{document}